\documentclass[11pt,a4paper,english]{article}      % Specifies the document class
\usepackage[english]{babel}
\usepackage{multicol}
\usepackage[T1]{fontenc}
\usepackage[latin1]{inputenc}
\usepackage[]{graphicx}
\usepackage[]{graphics}
\usepackage[]{latexsym}
\usepackage[pdftex]{hyperref}
\usepackage{geometry}
\title{\textbf{Stability Analysis of Possible Terrestrial Planets\\ in the HD 60532 System}}  % Declares the document's title.
\author{S. Rothwangl, S. Eggl and R. Dvorak}      % Declares the author's name.
\date{\footnotesize{Institute for Astronomy (IfA), University of Vienna,
              T\"urkenschanzstra\ss{}e 17, A-1180 Vienna\\}}      % Deleting this command produces today's date.

\begin{document}             % End of preamble and beginning of text.
\renewcommand{\refname}{\large{References}} 

\maketitle
%\thispagestyle{empty}                   % Produces the title.
%\newpage
%\tableofcontents
%\np

%\twocolumn

%\noindent

\noindent
\textbf{Abstract:} Investigations were carried out on the stability of the two gas giant system HD 60532 itself including additional, smaller bodies. The methods used were numerical simulations with the Lie Series integration method, and secular perturbation theory. The aims were to find stable regions for potential Earth-like planets within the system, and confirming the stability of the detected planets. As results, the given configuration can be considered stable for a period of at least 65 million years. The two planets create a large unstable section between 0.25 and 3.6 AU, where no additional body is able to remain stable. Variations in the mutual inclination between test bodies and the rest of the system of up to 20$^\circ$\ have little effects on the size of the zone of instability.

\indent

\begin{multicols}{2}

\section{\large{Introduction}}

Discovered in 2008 during a radial velocity survey of A to F type stars with the HARPS spectrograph at La Silla observatory, HD 60532 is a planetary system with two Jupiter-sized planets of minimum masses of 1.03 $\pm$ 0.05 and 2.46 $\pm$ 0.09 $M_J$ orbiting an F-type star (\hyperref[harps2]{Desort et al. 2008}). 
The planets are located at 0.76 and 1.58 AU respectively, suggesting a $3:1$ resonance, 
a claim which was supported by the work of Laskar and Correia (\hyperref[laskar]{2009}). 
The authors investigated the overall stability of the system as well as the most likely inclination of the discovered planets to the line of sight. 
First numerical analysis of the stability of additional bodies for two inclinations of the massive planets to the line of sight of the observations ($90^\circ$\ and $20^\circ$) were performed, identifying a zone of instability between 0.1 and 6 AU for $i=90^\circ$ and from 0.2 to 4 AU for $i=20^\circ$, thus suggesting an inclination of the two planets of about $20^\circ$ to the line of sight, corresponding to masses of 3.15 and 7.46 $M_J$. Periods of the planets were determined to be 201.46 $\pm$ 0.13 and 605.28 $\pm$ 2.12 days.
Further research on the formation of the system was published by Zs. S\'{a}ndor and W. Kley (\hyperref[formation]{2010}).\\

The article at hand shows the results of further investigations on the stability of the two discovered planets using numerical simulations and analytical investigations of secular perturbation effects.
Thereby we focus on the identification of stable zones for additional bodies which may have different inclinations with regard to the system.

%__________________________________________________________________

\section{\large{Methods}}
 
Numerical simulations of the system were performed by Lie integration as developed by Hanslmeier and Dvorak (\hyperref[hanslmeier]{Hanslmeier and Dvorak 1984}, Dvorak et al. \hyperref[chaos]{2005}). Tests on the stability of the two discovered massive planets were conducted for a time span of 10 to 65 million years. The length of integration was chosen in order to observe the full effects of secular perturbations and dominant resonances on the planets. Initial conditions for the  test calculations were the mean values of the planetary parameters given by Laskar and Correia (\hyperref[laskar]{2009}).\\
Secular resonance investigations for both the the planets and inserted massless bodies were calculated based on the derivations by Murray and Dermott (\hyperref[murray]{1999}).

%__________________________________________________________________

\section{\large{Results}}

\subsection{\small{Stability of the system HD 60532}}
Both planets show strong mutual interaction, as the orbits cover an area of 0.02 AU for planet b and 0.04 AU for planet c. Planetary eccentricities will vary between 0.00 to 0.19 for the outer planet and 0.03 to 0.37 for the inner planet. In the present configuration however, the system is stable. 

Variations for all elements except the semi-major axis increase in strength, if the inclination of both planets is not considered to be coplanar. A coupling of inclination and eccentricity of the inner planet is visible above 39.2 degrees mutual inclination, which can be attributed to the Kozai resonance (Kozai \hyperref[kozai]{1962}).

If both planets are positioned in the same plane or have only a small mutual inclination of one degree, both planets' arguments of perihelion rotate, while the longitudes of their ascending nodes librate in most cases, however, rotation also happens in one test run after about 7 million years when the initial inclination of both planets differs.

Using secular perturbation theory on the most stable configuration, as given in Dvorak et al. (\hyperref[chaos]{2005}), shows similar results as the numerical simulations. The mutual inclination, when starting with a difference of one degree, behaves almost exactly in the same way, with a period of about 750 years. Eccentricity, of planet b varies a little less, ranging between 0.22 and 0.28, and of c a little more, ranging between 0.01 and 0.1, with a period of about 1000 years. The differences in eccentricity between secular perturbation theory and numerical simulations may be explained by a possible $3:1$ mean motion resonance (MMR) of the gas giants. The longitude of pericenter rotates as well, the longitude of the ascending node librates.
%______________________________________________________________

\subsection{\small{Stability of additional bodies}}
The tests then were expanded to include additional massless bodies, since for planets of super Jupiter masses secular perturbations of planets of the size of Earth are assumed to be negligible.

\begin{figure}[h]
  \begin{center}$
    \begin{array}{cc}
      \includegraphics[width=5cm]{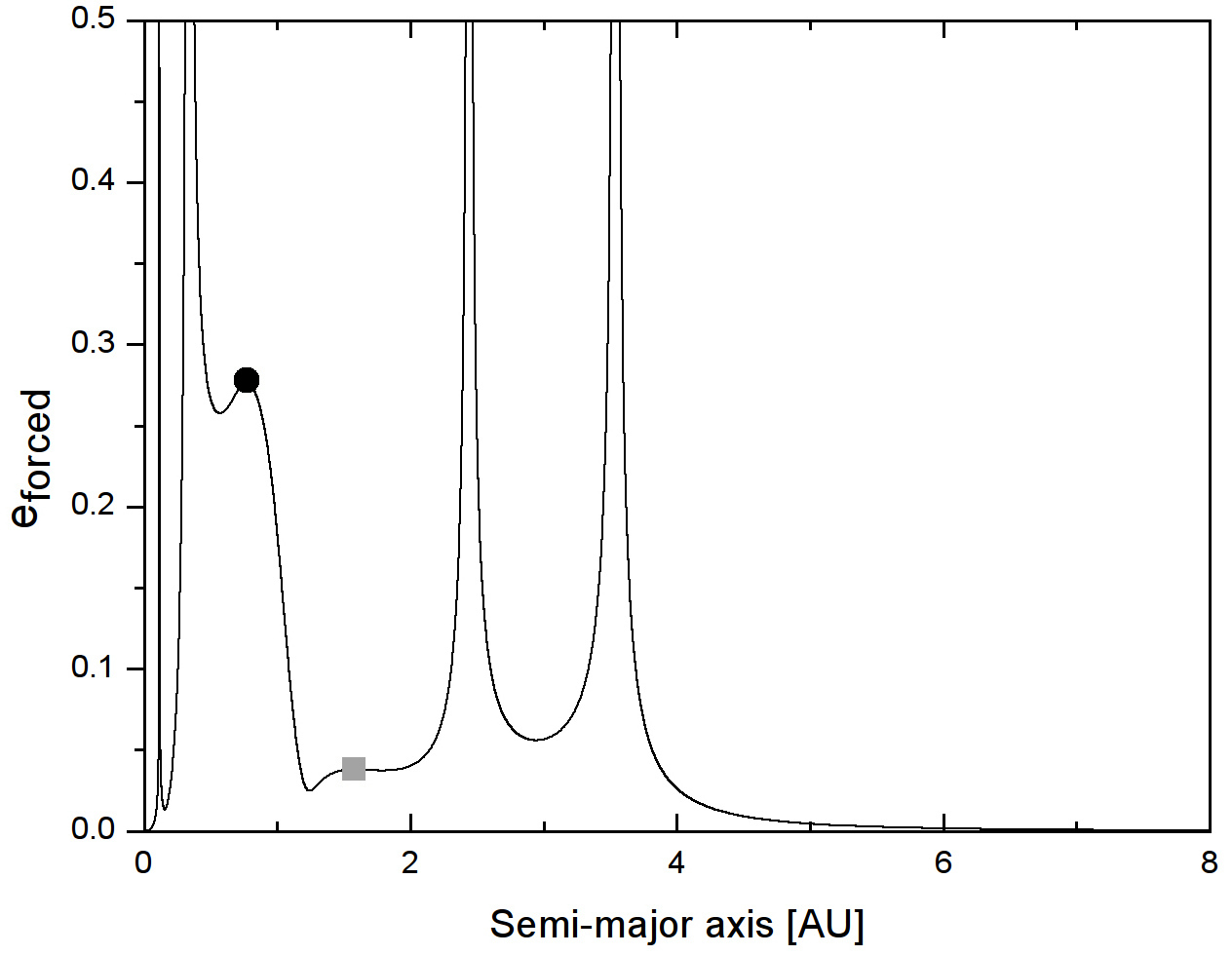} &
      \includegraphics[width=5cm]{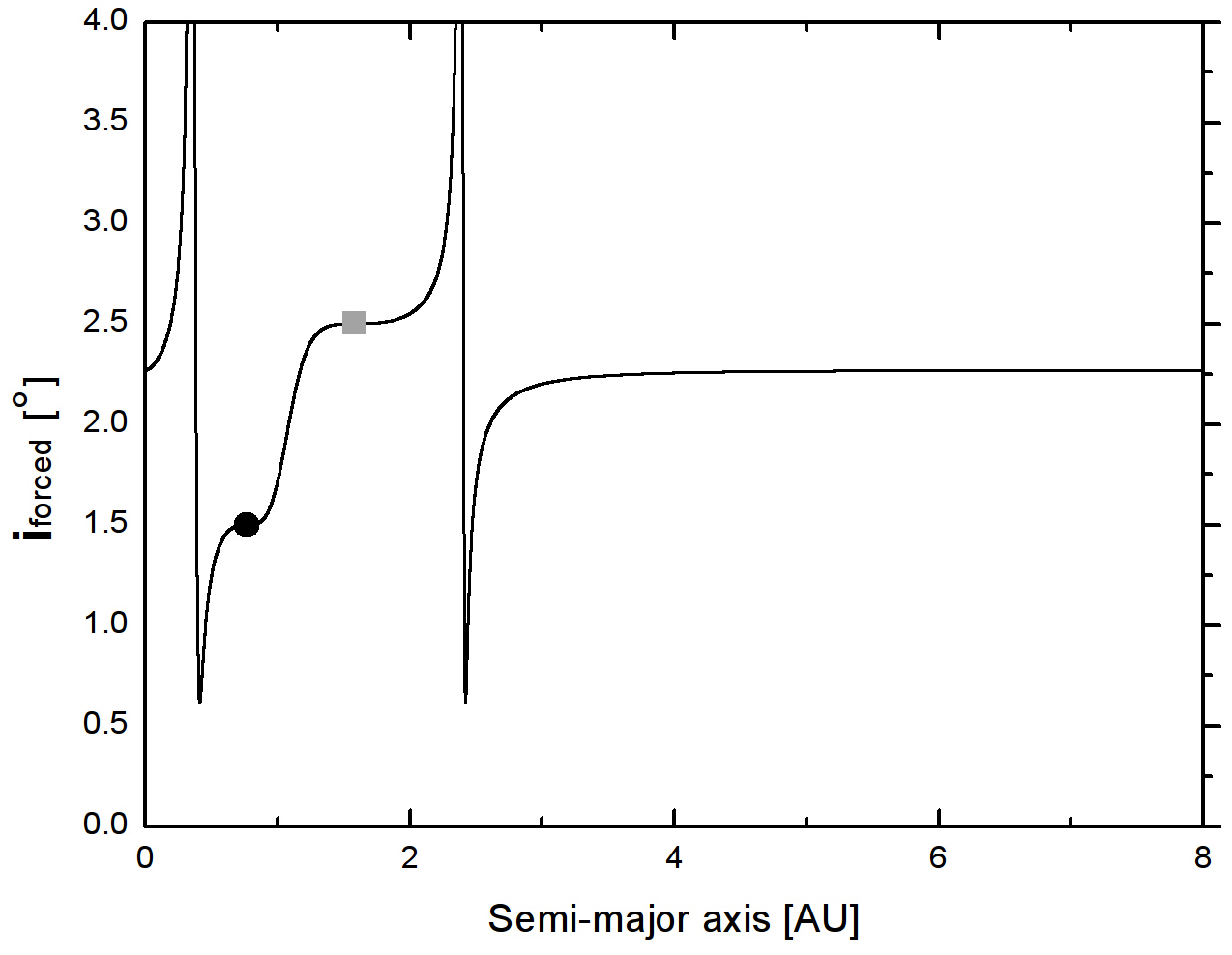}\\
      \includegraphics[width=5cm]{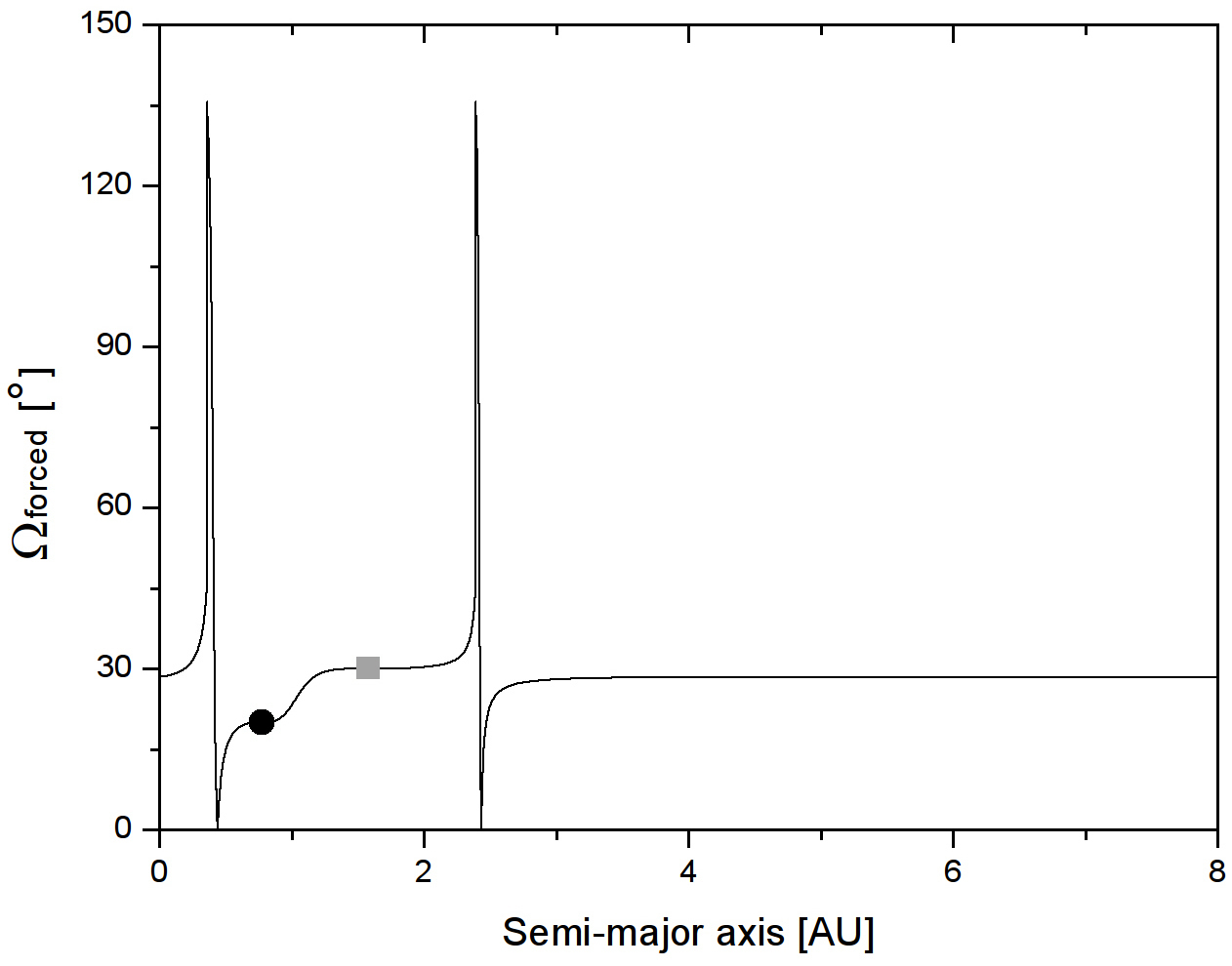} &
      \includegraphics[width=5cm]{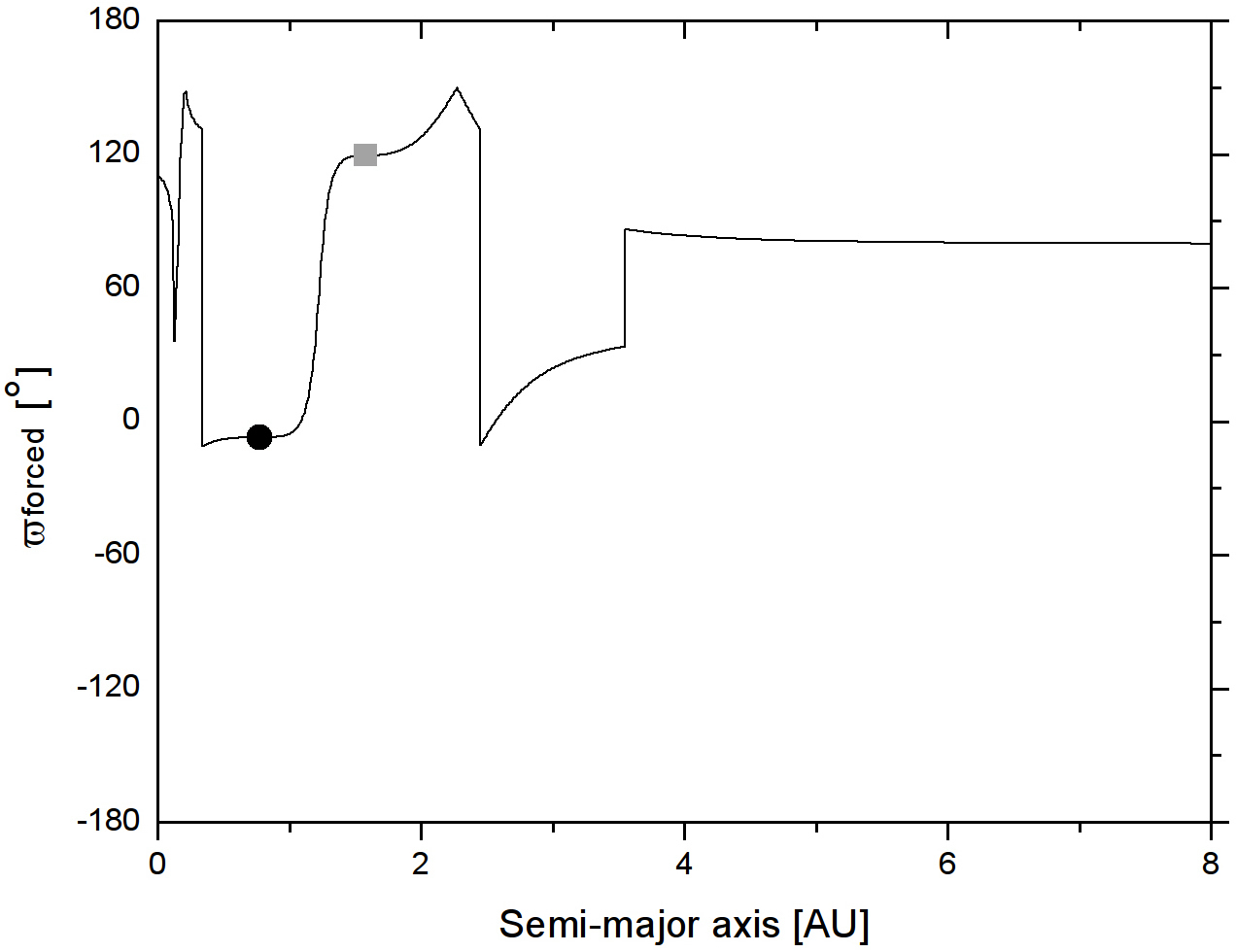} \\
    \end{array}$
    \caption{Forced eccentricity (top left), inclination (top right), longitude of the ascending node (bottom left) and perihelion (bottom right) of test bodies in the HD 60532 system. The black circle indicates the position of planet b, the grey box that of planet c.}
    \label{eforced}
  \end{center}
\end{figure}

Secular resonance investigations for massless bodies show that there exist several areas of instability. Eccentricity is elevated at 0.117 AU, although confined to a small space of less than 0.05 AU. A similar behavior can be seen for planetary semi-major axes around 0.3 AU
as well as between 2 and 4 AU. This can be seen in Figure \ref{eforced}. The black and grey dots in the graphs indicate the positions of HD 60532 b and c. Additionally, eccentricity is elevated around inner planet.

Between both planets and only little outside planet c, zones with only moderate excitement would exist, i.e. at 1.2 to 2.1 AU and at 3 AU. However even in these zones forced inclinations and eccentricities are elevated. Outside 5 AU, forced eccentricity is down to almost zero. Forced inclination is elevated to almost the value of planet c. The same is given for longitude of the ascending node and of perihelion of the test bodies.

\end{multicols}

\newpage
\begin{multicols}{2}

The results obtained via secular perturbation theory where then verified using numerical integration. Maximum integration for the test particles time was $10^6$ years, corresponding to up to 1000 secular periods of the eccentricity.

Four different inclinations were considered in the tests, i.e. 0$^\circ$, 5$^\circ$, 20$^\circ$\ and 25$^\circ$\ with respect to the orbital plane of the planets, which were assumed to be coplanar. The assumption that planets with orbits inclined to more massive planets can exist in the same system has been investigated by Lee and Thommes (\hyperref[lee]{2009}), however the particular rather lower values for the inclination were chosen since less massive planets generally experience an attraction towards the plane of the more massive planets. The test bodies' and massive planets' orbital parameters, as given by Laskar and Correia (\hyperref[laskar]{2009}), were:

\begin{center}
\begin{tabular}{|r|l|}\hline
\multicolumn{2}{|c|}{\textbf{Test particle starting conditions}\parbox[0pt][1.5em][c]{0cm}{}}\\
\hline
\textit{Semi-major axis} [AU] & 0.1 - 6.0\\
\textit{Eccentricity} & 0.0 - 0.1 \\
                      &(step size 0.1)\\
\textit{Inclination} [$^\circ$] & 0, 5, 20, 25\\
\textit{Argument of the perihelion} [$^\circ$] & 0\\
\textit{Longitude of the asc. node} [$^\circ$] & 0\\\hline
\end{tabular}
\end{center}

\begin{center}
\begin{tabular}{|r|c|c|}\hline
\multicolumn{3}{|c|}{\textbf{Massive planets starting conditions}\parbox[0pt][1.5em][c]{0cm}{}}\\
\hline
\textit{Planet}& b & c\\
\textit{Semi-major axis} [AU] & 0.7606 & 1.5854\\
\textit{Eccentricity} & 0.278 & 0.038\\
\textit{Inclination} [$^\circ$] & 0 & 0\\
\textit{Argument of the perihelion} [$^\circ$] & 352.83 & 119.49 \\
\textit{Longitude of the asc. node} [$^\circ$] & 0 & 0\\\hline
\end{tabular}
\end{center}

In order to investigate possible MMRs between test particles and gas giants, the initial planetary mean values were used, and test particles with an inclination of 0$^\circ$, semi-major axes of up to 14.0 AU were calculated. The longitude of the ascending node was also calculated with other initial conditions, however differences were negligible. Additionally, the starting conditions of the forced elements of the secular perturbation theory were tested for the area around 3 AU.

As no Kozai states were considered for the test particles, no major differences in their behavior was visible due to the variation of initial inclinations. Stable regions occur at 0.0 to 0.25 AU and 3.6 AU outwards, and a large unstable zone in between, where particles are perturbed heavily by the massive planets. Stability is possible inside respectively outside this area, where stability increases with increasing distance. Thus, the unstable zone includes the ones given by secular perturbation. No certain trend is given depending on the eccentricity of the test bodies in the innermost stable zone, while on the outside, planets with a low or average eccentricity, i.e. up to 0.06, have an increasing inward stability. The inner boundary of 0.25 AU, is thereby a sharper limit than the outer one, with test planets being unstable in less than 300 000 years at 0.27 AU and in most cases much faster. Escape time drops very fast outwards of this point towards the massive planets. Also, most of the test particles between 0.2 and 0.25 AU are ejected as well. All test planets remain stable at 0.18 AU. This again corresponds to secular results. Note also that slight trend for higher stability is given for higher inclinations. The outermost planet remaining stable have initial semi-major axis of 0.25 AU, eccentricities of 0.01, 0.02 and 0.05, and an inclination of 25$^\circ$. 

While all configurations of test particles with semi-major axes smaller than 0.18 AU were stable in the numerical simulations, especially independent of inclination and eccentricity, the value of 3.6 AU on the outer side can only be seen as boundary to which the first configurations remained stable. These configurations have an initial eccentricity 0.03 to 0.07 for 0$^\circ$\ inclination; an eccentricity of 0.02, 0.03 and 0.05 for 5$^\circ$\ inclination. The first configurations for initial inclinations of 20$^\circ$ and 25$^\circ$ were stable at 3.7 AU; configurations with 20$^\circ$\ had an initial eccentricity of 0.07 and 0.09, 25$^\circ$\ an eccentricity of 0.05 and 0.06.

Interestingly, a stable zone occurs at 3.7 and 4.0 AU, whereas at 4.0 AU, most configurations are again unstable. A possible explanation for this behavior could be that the $1:4$ resonance with massive planet c is very close to this value of semi-major axis. General stability is given outwards of 4.1 AU for all inclinations except 20$^\circ$\, which has unstable configurations at 4.3 AU. Note that according to calculations as given in Kasting et al. (\hyperref[kasting]{1993}) the habitable zone for this type of star is between 1.25 and 2.55 AU, and therefore being positioned in the unstable zone.

\begin{figure}[!ht]
   \begin{center}$
    \begin{array}{cc}
   \includegraphics[width=8.5 cm]{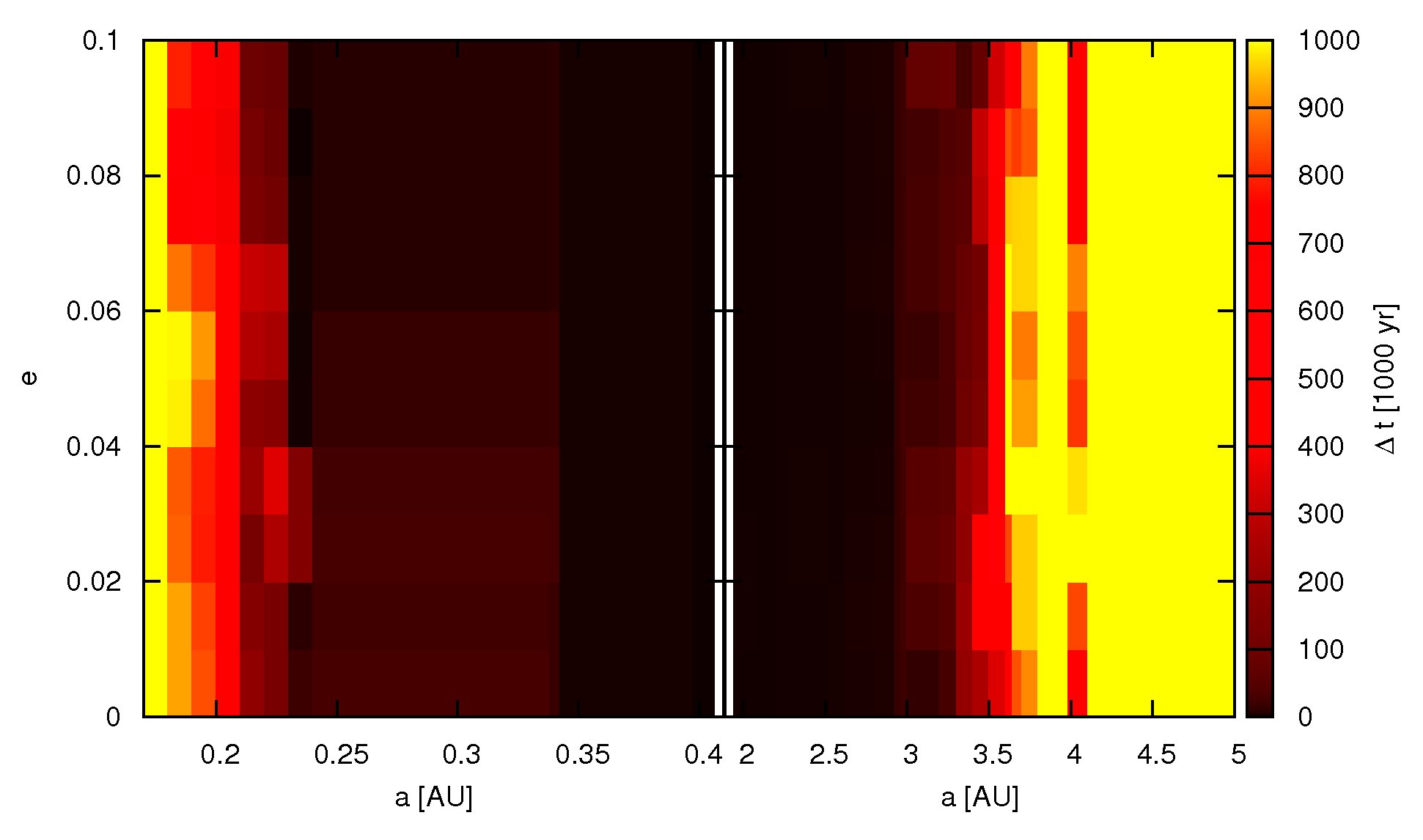} &
   \includegraphics[width=8.5 cm]{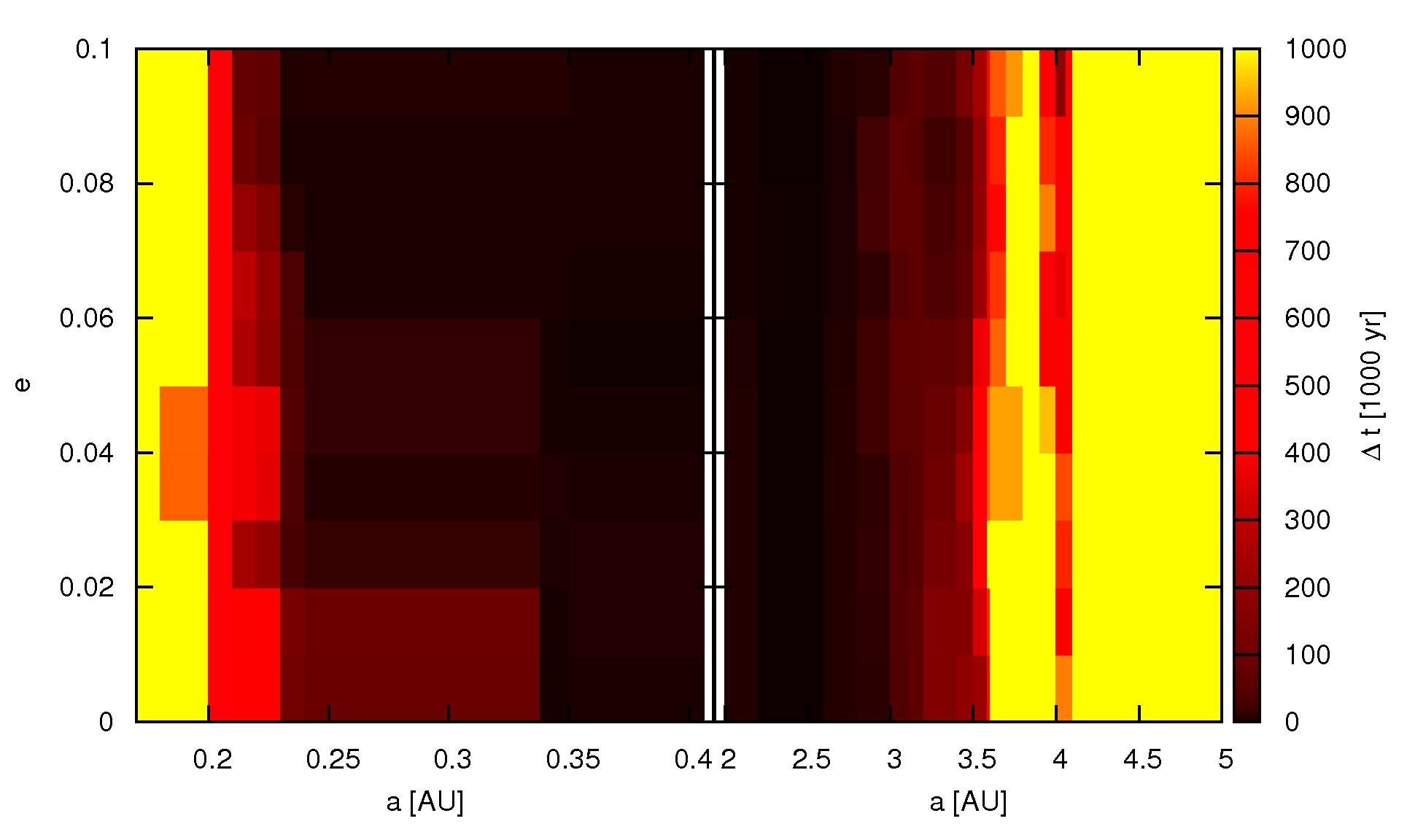} \\
   \includegraphics[width=8.5 cm]{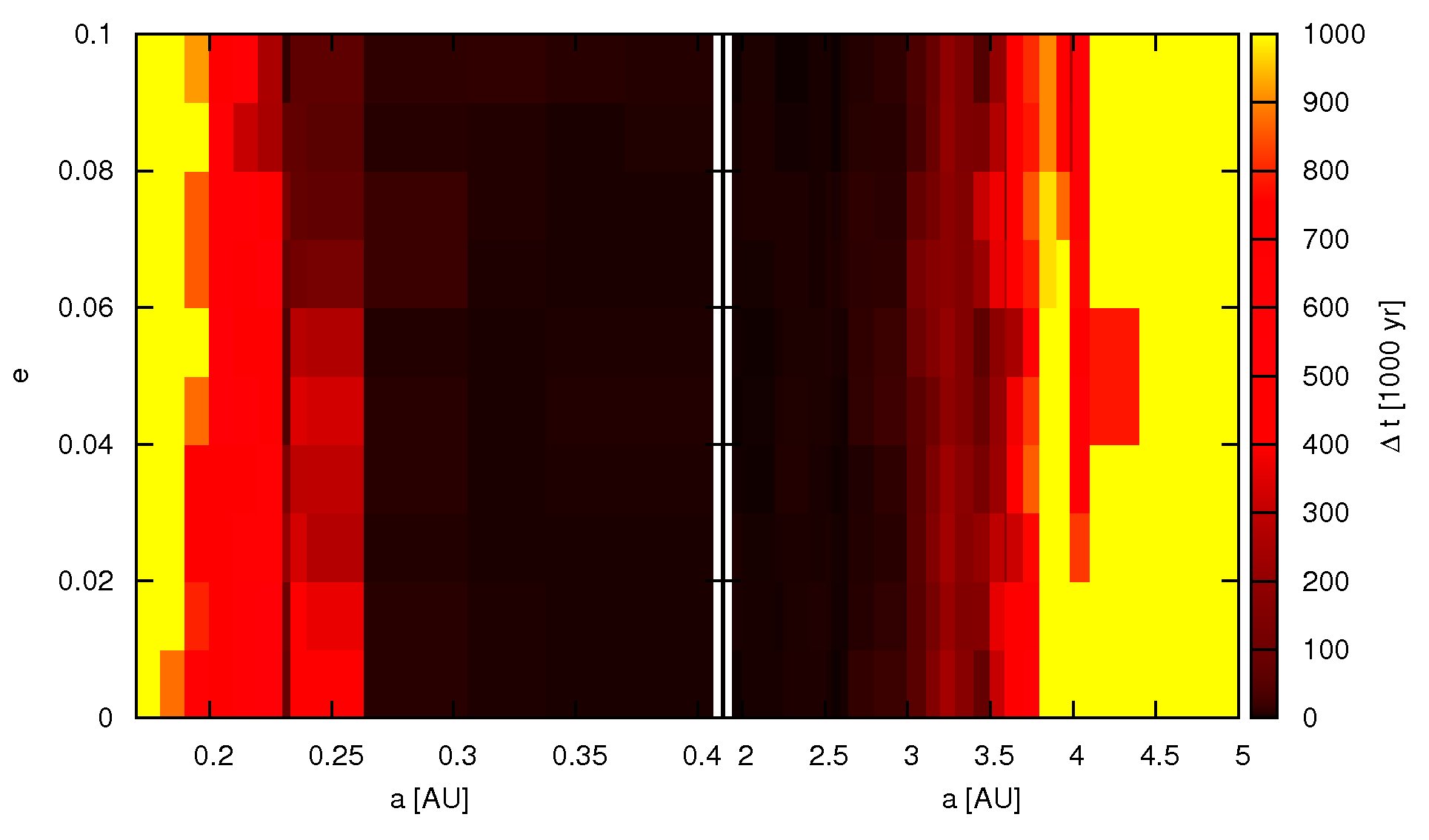} &
   \includegraphics[width=8.5 cm]{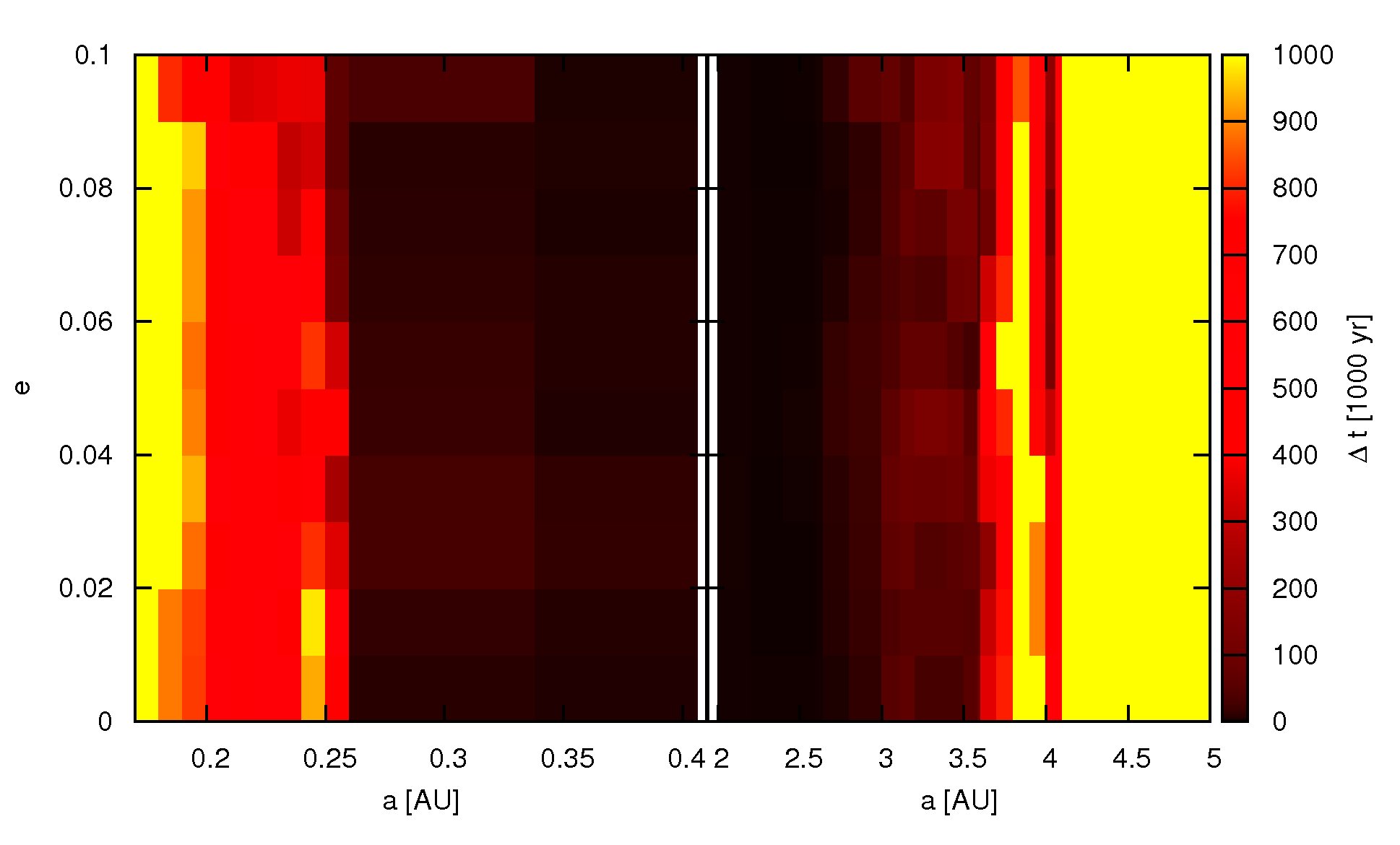}\\
    \end{array}$
    \end{center}
      \caption{Stability of test bodies with 0$^\circ$\ inclination (top left), 5$^\circ$\ inclination (top right), 20$^\circ$\ inclination (bottom left) and 25$^\circ$\ inclination (bottom right). Note that the middle section is not given completely, since all test bodies are unstable, and that the granularity of the left side differs from that of the right side. The colour coding is from yellow (very stable) through red to black (highly unstable).}
         \label{planets}
   \end{figure}

No unexpected behavior could be found for bodies in mean motion resonances other than $1:4$, as results were only stable if in an already stable region. Investigations included $1:1$, as well as higher and lower resonances. Consequently, both the Lagrangian points L4 and L5 of both planets are unstable. Ejection thereby happened very fast, with all bodies being ejected in less than 1000 years. This can be seen in Fig. \ref{planets}, picturing the stability of bodies with an inclination of zero degrees (top left); with an inclination of five degrees (top right);  with an inclination of twenty degrees (bottom left); and bodies with twenty-five degrees inclination (bottom right) with respect to the plane of the planets. The colour coding in these figures represents the escape time, from yellow (test planets that remained in the simulation for the whole integration time) through red to black (highly unstable with a short escape time of down to 1000 years), given in units of 1000 years. Planets with an eccentricity greater than 0.3 were thereby considered unstable. Also, the  semi-major axis was chosen as a criteria for stability because according to Murray and Dermott (\hyperref[murray]{1999}) the time derivative of the secular solution is zero. Thus, strong variations can be used as an indicator for unstable configurations.

Interestingly, the zone around 3 AU with only moderate forced eccentricity in the analytical investigations proved to be unstable in numerical integrations, even with the forced parameters of the secular perturbation theory as initial conditions. This will be the topic of further research. Apart from this, behavior of the eccentricity given by secular resonance is represented well in the stable zone of the numerical simulations. The average value of the inclination also coincides with the secular resonance calculations, however the amplitude of the oscillations can be as high as 40$^\circ$ on the boundary to the unstable zones. Note also that since the longitude of periapsis of the massive planets rotates in all our test cases, the secular excitation only plays a subordinate role in this setup.

\end{multicols}

\newpage

\begin{multicols}{2}
\section{\large{Conclusions}}
This article focuses on the identification of stable zones for additional bodies with different inclinations with respect to the orbital plane of the detected planets in the HD 60532 system. The system itself, with its massive planets, can be considered stable, nevertheless, these discovered planets show strong mutual interactions. Lower mass planets with stable orbits could exist inside ($a < 0.25$ AU) and outside ($a > 3.6$ AU) of the two already discovered planets of the HD 60532 system. Especially on the inside of the unstable zone that is reaching from 0.25 to 3.6 AU, different orbital parameters, as long as moderately given, have only little effect on the stability of the inserted hypothetical planet. The outer boundary proved to be a more dependent on the initial parameters, however still no big differences are shown.\\

%__________________________________________________________________

\noindent
\small{\textbf{Acknowledgements}}\\
 S. Rothwangl appreciates the financial support provided by a merit scholarship of the University of Vienna, and the help provided by B. Eckmair, I. Ru\ss{}mann and C. Saulder during the research and writing of this paper.
 S. Eggl acknowledges the support of University of Vienna's Forschungsstipendium 2012. The authors would like to acknowledge the support of the Austrian FWF.

\end{multicols}

\end{document}